\documentclass[a4paper]{jpconf}
\usepackage{graphicx}
\usepackage{amsmath,amssymb}

\usepackage[hidelinks,colorlinks]{hyperref}

\newcommand{\Neff}{N_\mathrm{eff}}
\newcommand{\Tcm}{T_\mathrm{cm}}
\newcommand{\vrho}{\varrho}
\newcommand{\bvrho}{\bar{\varrho}}

\begin{document}

\definecolor{carmine}{rgb}{0.59,0.,0.09}
\hypersetup{linkcolor=carmine,urlcolor=carmine,citecolor=carmine,urlcolor=carmine}

\title{Precision calculation of neutrino evolution in the early Universe}

\author{Julien Froustey}

\address{Institut d'Astrophysique de Paris, CNRS UMR 7095, Sorbonne Universit{\'e}, 98 bis Bd Arago, 75014 Paris, France}

\ead{julien.froustey@iap.fr}

\begin{abstract}
In the primordial Universe, neutrino decoupling occurs only slightly before electron-positron annihilations. This leads notably to an increased neutrino energy density compared to the standard instantaneous decoupling approximation, parametrized by the effective number of neutrino species $\Neff$. A precise calculation of neutrino evolution is needed to assess its consequences during the later cosmological stages, and requires to take into account multiple effects such as neutrino oscillations, which represents a genuine numerical challenge. Recently, several key improvements have allowed such a precise numerical calculation, leading to the new reference value $\Neff=3.0440$. 
\end{abstract}

\section{Introduction}

When the temperature of the Universe decreases below $2 \, \mathrm{MeV}$, neutrinos decouple from the plasma of photons, electrons and positrons, as the interaction rates drop below the Hubble expansion rate. Later, $e^+e^-$ pairs annihilate into photons, reheating essentially only the photons but not the neutrinos (when the temperature decreases below the electron mass $m_e = 0.511 \, \mathrm{MeV}$). 

If these events were well separated in time (so-called “instantaneous decoupling” approximation), then after decoupling the neutrino temperature would be $T_\nu = \Tcm \propto a^{-1}$ with $a$ the scale factor. Separate entropy conservation then leads, after $e^+e^-$ annihilations, to the standard ratio for the temperatures $T_\gamma / T_\nu = (11/4)^{1/3} \simeq 1.40102$.

However, neutrino decoupling takes place around $T \sim 1 \, \mathrm{MeV}$, hence this description is only approximate: neutrinos are partially reheated by $e^+e^-$ annihilations, a phenomenon called \emph{incomplete neutrino decoupling}. This leads to an increased (anti)neutrino energy density, historically parametrized by the effective number of neutrino species $\Neff$, defined such that long after decoupling
\begin{equation}
    \rho_\nu = \sum_{\alpha = e, \mu, \tau}{(\rho_{\nu_\alpha} + \rho_{\bar{\nu}_\alpha})} = \frac78 \left(\frac{4}{11}\right)^{4/3} \times \Neff \times \rho_\gamma \, , 
\end{equation}
with $\Neff=3$ in the instantaneous decoupling limit. Currently, the most stringent constraints on this parameter arise from Cosmic Microwave Background (CMB) experiments, the most recent values with 68 \% confidence levels being $\Neff = 2.99 \pm 0.17$ (\emph{Planck} + BAO \cite{Planck18}) and $\Neff = 3.13 \pm 0.30$ (SPT-3G 2018 + \emph{Planck} \cite{SPT-3G}).

This motivates the objective of providing a prediction of $\Neff$ in the Standard model, that is in the $\Lambda$CDM model of cosmology and the Standard Model of particle physics (adding the phenomenology of neutrino oscillations through the Pontecorvo-Maki-Nakagawa-Sakata (PMNS) matrix and the mass-squared differences), with a precision of a few $10^{-4}$. This requires to follow the evolution of neutrinos throughout the decoupling era with a sufficient accuracy, including all relevant physical ingredients.

A first step consists in neglecting flavour mixing, in which case one needs to solve a Boltzmann equation for neutrino distribution functions~\cite{Dolgov_NuPhB1997,Esposito_NuPhB2000,Grohs2015,Froustey2019}. It was then realized that, since particles do not evolve in the vacuum but in a thermal bath, QED corrections to the plasma thermodynamics needed to be taken into account. They were first included in a calculation of neutrino decoupling at order $\mathcal{O}(e^2)$ in~\cite{Mangano2002}. Finally, taking into account flavour mixing requires to promote the set of distribution functions to a full density matrix (see equation~\eqref{eq:vrho} below) and to introduce a corresponding generalization of the Boltzmann equation. This formalism was used to reach the value $\Neff \simeq 3.045$~\cite{Mangano2005,Relic2016_revisited}, with one caveat: the off-diagonal components of the collision term were evaluated with a damping approximation. Recently, it was shown in~\cite{Bennett2020} that it is necessary to include the QED corrections up to $\mathcal{O}(e^3)$ order to reach the desired accuracy on $\Neff$, since these corrections were expected to decrease $\Neff$ by $10^{-3}$---a result checked in~\cite{Akita2020}.

We present here a summary of the results obtained in~\cite{Froustey2020}, to which we refer for details on the numerical procedure and a detailed analysis of the physics of flavour oscillations in this setup.

\section{Quantum Kinetic Equations} \label{sec:QKE}

In order to account for flavour mixing, the statistical ensemble of neutrinos is described by a generalization of occupation numbers, that is by a density matrix which reads
\begin{equation}
\label{eq:vrho}
    \vrho(p,t) = \begin{pmatrix}
    f_{\nu_e} & \vrho_{e \mu} & \vrho_{e \tau} \\
    \vrho_{\mu e} & f_{\nu_\mu} & \vrho_{\mu \tau} \\
    \vrho_{\tau e} & \vrho_{\tau \mu} & f_{\nu_\tau}
    \end{pmatrix} \, ,
\end{equation}
in which the diagonal elements correspond to the standard distribution functions, while the off-diagonal components result from flavour mixing. In general one must define a similar quantity $\bvrho$ for antineutrinos, but in the case of vanishing asymmetry we consider here $\bvrho = \vrho$.

We define the effective\footnote{The decoupling process being an out-of-equilibrium one, (anti)neutrinos do not keep Fermi-Dirac spectra and there are no actual temperatures.} temperatures $T_{\nu_\alpha}$ and the non-thermal distortions $\delta g_{\nu_\alpha}$ through
\begin{equation}
    f_{\nu_\alpha}(p,t) = \frac{1}{e^{p/T_{\nu_\alpha}}+1}\left[1 + \delta g_{\nu_\alpha}(p,t)\right] \, , \qquad \text{where} \quad \rho_{\nu_\alpha} \equiv \frac78 \frac{\pi^2}{30} T_{\nu_\alpha}^4 \, .
\end{equation}
This allows to separate the energy density contribution (via $T_{\nu_\alpha}$) and the residual spectral distortions. In the following, I summarize results on the energy density since it is the relevant quantity to compute $\Neff$, but the distortions are key to understanding, for instance, the effect of incomplete neutrino decoupling on Big Bang Nucleosynthesis (BBN)~\cite{Froustey2019,Froustey2020}.

The evolution of $\vrho$ is given by the Quantum Kinetic Equation (QKE)~\cite{Froustey2020,SiglRaffelt,BlaschkeCirigliano,Volpe_2013}
\begin{equation}
\label{eq:QKE}
    i \left[\frac{\partial}{\partial t} - H p  \frac{\partial}{\partial p} \right] \vrho(p,t) = \left[ U \frac{\mathbb{M}^2}{2p} U^\dagger , \vrho \right] - 2 \sqrt{2} G_F p \left[ \frac{\mathbb{E}_e + \mathbb{P}_e}{m_W^2}, \vrho \right] + i \, \mathcal{C} \, .
\end{equation}
On the right-hand side of this equation:
\begin{itemize}
    \item the first term is the vacuum contribution, involving $\mathbb{M}^2$, matrix of mass-squared differences, and the PMNS matrix $U$ ;
    \item the second is the mean-field potential, with $\mathbb{E}_e = \mathrm{diag}(\rho_{e^-}+\rho_{e^+},0,0)$ and likewise for the pressure density ;
    \item $\mathcal{C}$ is the collision term, which accounts for scattering and annihilations with electrons and positrons, and also among (anti)neutrinos.
\end{itemize}
In addition to the QKEs~\eqref{eq:QKE} written for each momentum (and coupled through the collision term), and the corresponding equations for antineutrinos, the total energy conservation equation $\dot{\rho} + 3H(\rho + P)$ is rewritten as an equation on the photon temperature $T_\gamma$. QED corrections to the plasma thermodynamics~\cite{Bennett2020} enter through this equation.

\section{Results} \label{sec:Results}

The previous set of equations is solved with the numerical algorithm \texttt{NEVO}, which notably relies on a discretization of the momentum grid (through a Gauss-Laguerre quadrature) and a direct computation of the Jacobian of the differential system~\cite{Froustey2020}. 

\begin{figure}[!h]
\centering
\begin{minipage}{0.48\textwidth}
\includegraphics[width=\textwidth]{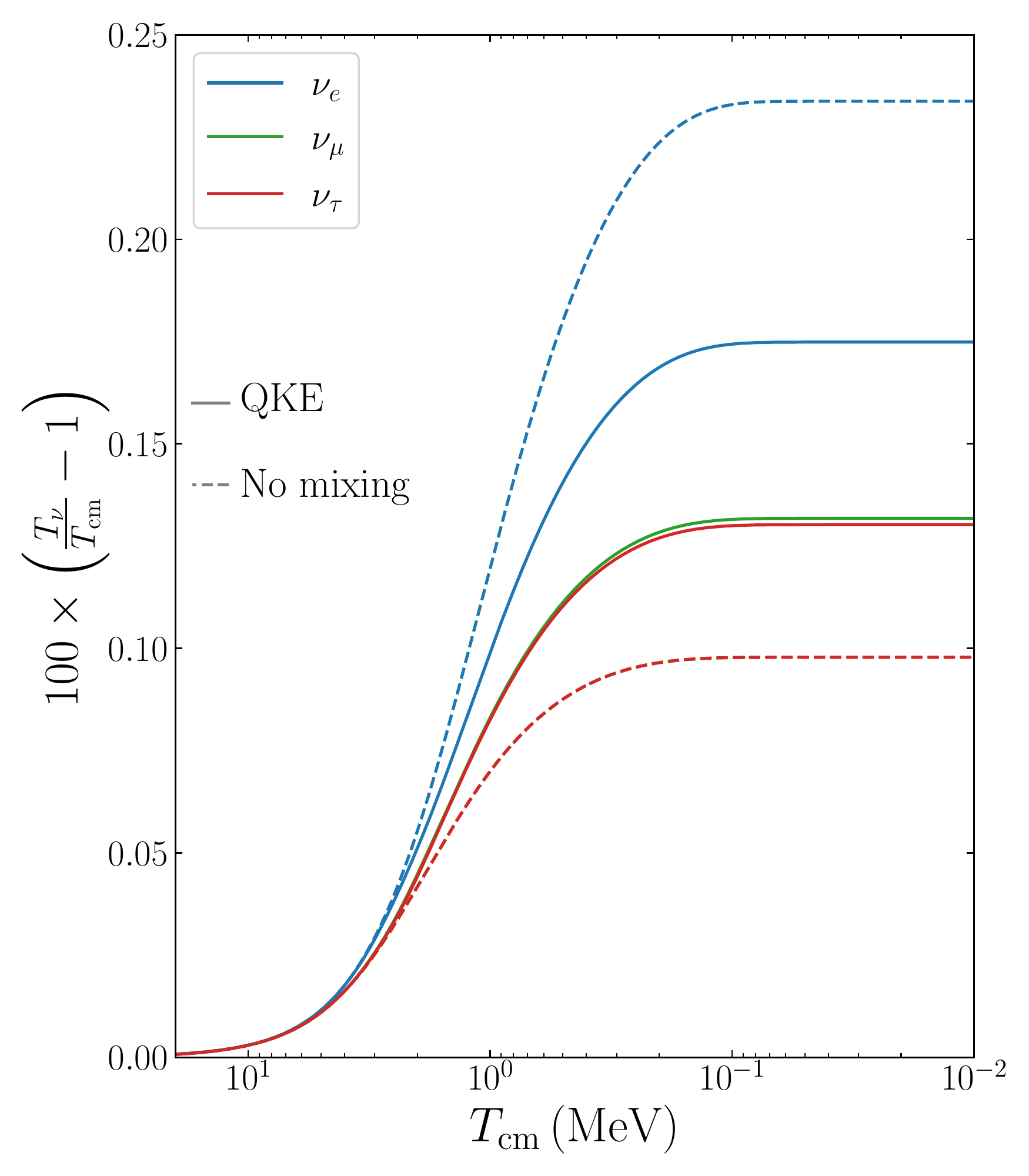}
\caption{\label{fig:results_Tnu} Evolution of the effective neutrino temperatures, with (solid lines) and without (dashed lines) flavour mixing.}
\end{minipage}\hspace{0.04\textwidth}%
\begin{minipage}{0.48\textwidth}
\includegraphics[width=\textwidth]{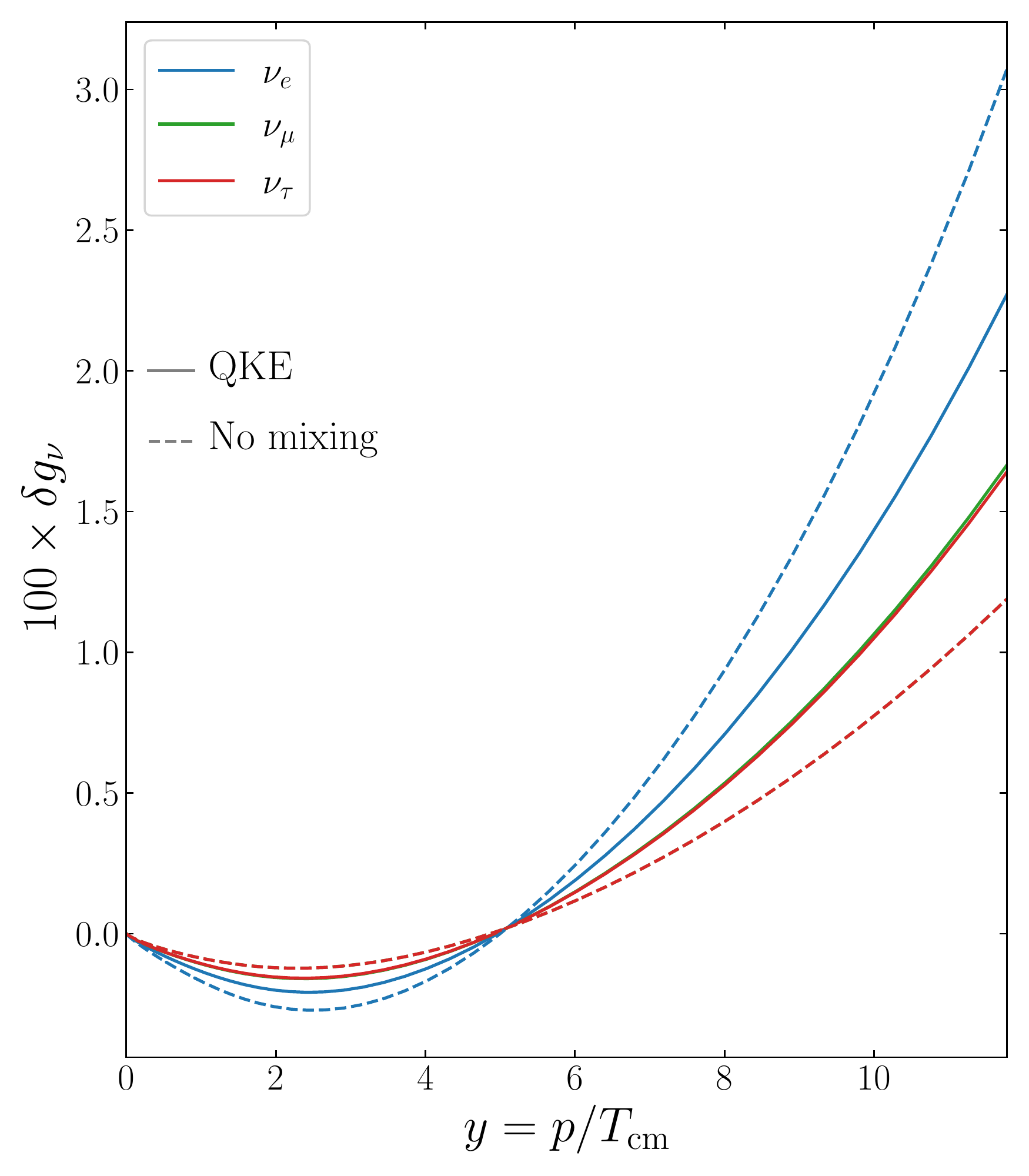}
\caption{\label{fig:results_gnu} Frozen-out effective spectral distortions (i.e.~for $\Tcm = 0.01 \, \mathrm{MeV}$), with and without flavour mixing.}
\end{minipage} 
\end{figure}

In figure~\ref{fig:results_Tnu} the evolution of the neutrino effective temperatures is depicted with and without flavour oscillations. Due to the existence of charged-current processes with $e^\pm$ (that do not exist for muon and tau neutrinos), the transfer of entropy from electrons and positrons is increased towards $\nu_e$, hence the higher values of $T_{\nu_e}$. Likewise, the non-thermal residual distortions are more important for $\nu_e$ (see figure~\ref{fig:results_gnu}).

Once decoupling is complete, all comoving quantities are frozen-out and the final values are reported in table~\ref{tab:results}. The instantaneous decoupling values are given as a reference, and they do not include any QED corrections. On the contrary, the no mixing and QKE cases include QED corrections to the plasma thermodynamics up to $\mathcal{O}(e^3)$. Flavour oscillations reduce the differences between the different flavours, a feature clearly seen on figures~\ref{fig:results_Tnu} and~\ref{fig:results_gnu}.

\begin{table}[!h]
\caption{\label{tab:results}Numerical values of the frozen-out effective temperatures.}
\begin{center}
\begin{tabular}{llllll}
\br
Final values & $T_\gamma/\Tcm$ & $T_{\nu_e}/\Tcm$ & $T_{\nu_\mu}/\Tcm$ & $T_{\nu_\tau}/\Tcm$ & $\Neff$ \\
\mr
Instantaneous decoupling & $1.40102$ & $1.00000$ & $1.00000$ & $1.00000$ & $3.0000$ \\
No mixing & $1.39800$ & $1.00234$ & $1.00098$ & $1.00098$ & $3.0434$\\
QKE & $1.39797$ & $1.00175$ & $1.00132$ & $1.00130$ & $3.0440$ \\
\br
\end{tabular}
\end{center}
\end{table}

The large separation of scales between the oscillation frequencies and the collision rate allows to average over these oscillations (which explains the absence of any oscillatory behaviour on figure~\ref{fig:results_Tnu}). It amounts to discarding the off-diagonal parts of the density matrix in the matter basis (i.e., the basis which diagonalizes the Hamiltonian [vacuum + mean-field]). This simplified scheme was named \emph{Adiabatic Transfer of Averaged Oscillations} (ATAO), and the results are indistinguishable from the QKE ones, while substantially reducing the computation time. Indeed, the shortest time scale is averaged upon and one only has to keep track of three quantities for each momentum (the diagonal entries of $\vrho$ in matter basis), contrary to the full $\vrho$ matrix in a QKE resolution~\cite{Froustey2020}.

\subsection*{Effective number of neutrinos}

As indicated in table~\ref{tab:results}, we get in the “full” case (that is, including in particular the full collision term and $\mathcal{O}(e^3)$ QED corrections):
\begin{equation}
\label{eq:Neff}
    \boxed{\Neff = 3.0440} \, ,
\end{equation}
with a precision of a few $10^{-4}$, due to the experimental uncertainty on the physical parameters (mostly the mixing angle $\theta_{12}$) and the numerical variability depending on the settings of our algorithm. This value was later confirmed in an independent work~\cite{Bennett:2020zkv} which also included all the physical ingredients mentioned above.

The value of $\Neff$ with mixing is slightly larger than its value without. Indeed, the mixing and mean-field terms tend to depopulate the more numerous $\nu_e$ and populate the other flavours, which frees some phase space for the reactions which create $\nu_e$, while increasing the effect of Pauli-blocking factors for reactions creating $\nu_{\mu, \tau}$. Since the former are the dominant reactions (thanks to charged-current processes), the net effect is a larger entropy transfer from $e^\pm$, hence the larger value of $\Neff$.

\subsection*{Neutrino energy density today}

The neutrino temperatures today ($T_\nu^0 \sim 1.7 \times 10^{-4} \, \mathrm{eV}$) being smaller than at least two of the neutrino masses (given the values of the mass-squared differences~\cite{PDG}), at least two neutrino mass eigenstates are non-relativistic today and have a non-negligible contribution to the energy density in the Universe, given by $\rho_{\nu_i} = m_{\nu_i} n_{\nu_i}$ for the eigenstate $\nu_i$. The number densities can be obtained from the results given in table~\ref{tab:results} (converted in the mass basis), which allows to revise the value previously obtained in~\cite{Mangano2005}. Assuming that all three eigenstates are non-relativistic today ($m_{\nu_1}\simeq m_{\nu_2} \simeq m_{\nu_3} \gg T_\nu^0$), we get:
\begin{equation}
    \boxed{\Omega_\nu = \frac{\rho_\nu^0}{\rho_\text{crit}^0} = \frac{\sum_{i}{m_{\nu_i}}}{93.12 \, \mathrm{eV} \times h^2}} \, ,
\end{equation}
with $h$ the current value of the Hubble parameter in units of $100 \, \mathrm{km \cdot s^{-1} \cdot Mpc^{-1}}$. Note that this expression also takes into account updated values for the physical constants such as $\mathcal{G}_N$~\cite{PDG}.

\section{Conclusion}

We have presented in~\cite{Froustey2020} a new calculation of neutrino decoupling in the early Universe, which included for the first time the full matrix structure of the collision integrals, in addition to the necessary QED corrections to reach a few $10^{-4}$ precision on $\Neff$. The results lead to the recommended value $\Neff = 3.0440$, and a revised value for today's neutrino energy density $\Omega_\nu$.

\ack

I thank Julien Lesgourgues for pointing out the question of today's neutrino energy density. 
The original work~\cite{Froustey2020} was done in collaboration with C.~Pitrou and M.~C.~Volpe.

\section*{References}
\bibliographystyle{iopart-num}
\bibliography{BiblioProceedings}

\end{document}